\documentclass[twocolumn,amsmath,amssymb]{revtex4}
\usepackage{graphicx}
\usepackage{hyperref}
\begin{document}

%\usepackage{graphicx}
%\usepackage{hyperref}
%\begin{document}

%\title{Observers and the Mathematical Ensemble}
\title{The Observer Class Hypothesis}

%\date{2007.02.25}
%\date{2008.02.26}
\date{\today}

\author{Travis Garrett}
%\email{tmgarret@physics.unc.edu}
%\email{garrett@phys.lsu.edu}
\email{tgarrett@perimeterinstitute.ca}
%\affiliation{UNC Chapel Hill}
\affiliation{Perimeter Institute for Theoretical Physics, Waterloo, Ontario N2L 2Y5, Canada}
\affiliation{Department of Physics \& Astronomy, Louisiana State University, Baton Rouge, LA 70802, USA}

\newcommand{\be}{\begin{equation}}
\newcommand{\ee}{\end{equation}}

\begin{abstract}   

The discovery of a small cosmological constant 
has stimulated interest in the measure problem.
One should expect to be a typical observer, 
but defining such a thing is difficult in the vastness of an eternally 
inflating universe.  We propose that a crucial 
prerequisite is understanding why one should 
exist as an observer at all.  We assume that the 
Physical Church Turing Thesis is correct and  
therefore all observers (and everything else that exists)
can be described as different types of information.
We then argue that the observers 
collectively form the largest class of information
(where, in analogy with the Faddeev Popov procedure, 
we only count over ``gauge invariant" forms of 
information).
The statistical predominance of the observers 
is due to their  
ability to selectively absorb other forms of information 
from many different sources. 
In particular, it is the combinatorics that arise from   
this selection process which  
leads us to equate the observer class $\mathcal{O}$ with the nontrivial 
power set $\hat{\mathcal{P}}(\mathcal{U})$ 
of the set of all information $\mathcal{U}$.
Observers themselves are thus the typical 
form of information. 
%This has interesting 
%consequences for the measure problem, 
%and leads to dramatic long 
%term predictions.
If correct, this proposal simplifies the measure problem, and leads 
to dramatic long term predictions.

\end{abstract}

\maketitle

\section{Introduction}

%\rmfamily
%\sffamily
%\ttfamily

%cmr & Computer Modern Roman 
%cmss & Computer Modern Sans 
%cmtt & Computer Modern Typewriter 
%cmm & Computer Modern Math Italic 
%cmsy & Computer Modern Math Symbols 
%cmex & Computer Modern Math Extensions 
%ptm & Adobe Times 
%phv & Adobe Helvetica 
%pcr & Adobe Courier

%\fontfamily {pcr}\selectfont

%Let's test bib-stuff \cite{aguirre05,bekenstein73,bojowald03,carroll00}
%As opposed to: \cite{chen04}\cite{dyson79}\cite{freivogel04}

% hyperlinks in the text:
%\href{http://www.arxiv.org}{arxiv}

The detection of ongoing cosmological acceleration 
was one of the most interesting discoveries in recent years
\cite{Riess:1998cb},\cite{Perlmutter:1998np}.
The form of the acceleration (with $w$ close to $-1$)
leads to the conclusion that it is probably 
due to a small cosmological constant 
$\Lambda$ \cite{Weinberg:1988cp},\cite{Trodden:2004st},\cite{Nobbenhuis:2004wn}.
Einstein's ``biggest blunder" has turned out not to 
be a blunder at all.

In a further amusing twist some
explanations for the source of this $\Lambda$ involve arguments 
about typical observers within a vast multiverse.
A large amount of fine tuning is needed to reduce 
$\Lambda$ from its natural value of $M^4_P$,
and this fine tuning winds up at a number very close to the 
anthropic bound, confirming Weinberg's 
prediction \cite{Weinberg:1987dv}.  
Furthermore string theory provides a wide enough 
landscape of states to generate such a 
small $\Lambda$ 
\cite{Bousso:2000xa,Feng:2000if,Kachru:2003aw,Susskind:2003kw}, 
and eternal inflation 
gives a method to populate them \cite{Linde:1986fd},\cite{Aguirre:2007gy}.
These anthropic arguments 
\cite{Vilenkin:1994ua,Tegmark:1997in,Martel:1997vi,Garriga:1999bf,Carter:2006gy} thus reintroduce 
the somewhat unusual subject matter of observers in physics, 
some 80 years after 
the discovery of modern quantum mechanics
in the late 1920s.

Indeed, one should expect to exist as a ``typical observer": 
one shouldn't see $50 \sigma$ events or 
macroscopic violations of the 2nd law and so forth.
This is a reflection of the statistical nature of 
our universe, and practically speaking it is 
implicitly assumed in all scientific disciplines.
The issue of what constitutes a typical observer 
becomes somewhat nontrivial however when 
considering the immense volumes of spacetime 
generated in cosmology.

%

%\cite{hartle07},\cite{srednicki10}
  
For instance, upon making 
some initially reasonable-sounding 
assumptions 
one could come to the 
conclusion that we exist in the earliest 
civilization possible after the big bang 
(the youngness paradox \cite{Guth:2007ng},\cite{Bousso:2007nd}),
or that the typical observer is a Boltzmann Brain 
(i.e. a disembodied 
brain floating in the void \cite{Linde:2006nw},\cite{DeSimone:2008if}).
Trying to fix these bizarre results 
composes the measure problem:
how to define probabilities 
in the infinite volumes generated 
by eternal inflation 
%\cite{Weinberg:1988cp},\cite{Weinberg:2000yb},\cite{Weinberg:2005fh},
%\cite{Freivogel:2005vv},\cite{Polchinski:2006gy},\cite{Susskind:2007pv},
%\cite{Wilczek:2005aj},\cite{Hartle:2007zv},\cite{Srednicki:2009vb},
%\cite{Linde:2010xz},\cite{Linde:2009ah}.  
\cite{Weinberg:2000yb,Weinberg:2005fh,Tegmark:2005dy,
Freivogel:2005vv,Banks:2005ru,Aguirre:2006ap,Polchinski:2006gy,Susskind:2007pv,
Wilczek:2005aj,Page:2006ys,Hartle:2007zv,Srednicki:2009vb,
Page:2008mx,Linde:2010xz,Linde:2009ah,Aguirre:2010rw}.  
A proposed measure solution 
will posit some form of cutoff, so that only a finite 
number of observers are considered,
with the statistical distribution of their 
experiences (very broadly speaking) thus well defined.
There are measures that just consider the 
observations made along a single 
world line, and others that sample from 
entire spacetime volumes.
The hope is that with the correct choice of measure 
the standard, non-exotic experiences of observers within 
our Hubble volume will be found to be 
typical within the entire multiverse, although this is 
still an open question.

There is a very deep and interesting assumption 
behind all of these ideas, one which is 
generally not questioned.  
As noted, one should expect 
to exist as a typical observer.
However, why should one expect to exist as an 
observer at all?
This is a nonintuitive subject matter (and it is something 
that is quite easy to take for granted), so 
it is helpful to take some time and 
rephrase the question in a number of 
different ways.

It is important to first clarify: what are observers? 
We staunchly support the  
materialist viewpoint and posit that the 
brain/mind connection is complete: all of one's
conscious experiences as an observer are 
generated by the interactions of neurons in one's brain
(see e.g. \cite{Koch:2004},\cite{Baum:2006},
\cite{Hawkins:2005},\cite{Damasio:2000}).
In turn, observers exist due to natural selection: 
being able to extract pertinent information from 
one's environment is a very useful 
evolutionary adaptation.

To give a more detailed example, 
consider an 
individual having breakfast in the morning.
At one moment in time they will experience a
particular mixture of sights and sounds -- the sunlight 
coming through the window, the coffee brewing --
perhaps while having a thought about the upcoming day 
or remembering something from the previous one,
and all of this is permeated by an emotional milieu and a sense of self 
in the background.
All of these aspects of consciousness are equal to
the patterns created by the 
subset of neurons that are active 
in the individual's brain at that moment in time.

To transition to a more abstract description, 
the experiences of that individual are a particular type of pattern:
a form of topological connected graph, 
as realized by the connections
of the currently interacting neurons.
The precise structure of the pattern 
(and thus the nature of the conscious experiences)
will change from moment to moment, based on the 
dynamics of the neural biophysics and the 
details of the incoming sensory information.
%\footnote{Note that this is a ``mesoscale" description -- 
%one could also use a ``macroscale" description: 
%the individual's experiences change based on what 
%they choose to think about and do, 
%or a ``microscale" description: the patterns change 
%based on the laws of atomic physics.  All of these 
%are valid and just differ in the number of details 
%kept.}.
There is a large amount of self-similarity however in the structure 
of the individual's neural patterns as they evolve in time, 
and indeed there is substantial similarity between different people's 
conscious states (e.g. we can talk to each other). 
%\footnote{Based on anatomical structure, 
%sand for instance the fact that we can talk to each other.}.
Collectively all of these types of patterns form a 
class of mental states.  To rephrase the typical 
observer assumption, one expects to exist as an element 
of this class, where each member is 
a type of pattern 
formed out of an arrangement of atoms.

%The neurons themselves are all composed 
%out of atoms, so to abstract away even further, the 
%observer is a particular type of pattern created 
%out of an arrangement of atoms.
%To reiterate, most of the observers 
%in our Hubble volume are typical 
%(e.g. they see the 2nd law is obeyed), 
%so given that one exists as an observer, 
%one should furthermore exist as a typical observer. 
%But why should one exist as an observer -- 
%as a member of a particular class of patterns formed out of atoms -- at all?

%Before proposing an answer, let's give an even 
%more abstract definition of the problem.

Further abstraction is possible and is useful. 
The key is that observers are just a particular 
type of information, as is everything else.
%\footnote{Here we will quickly and practically 
%define information to be that 
%which a computer processes, although we will 
%refine this somewhat going forward.}.
That is, we assume that the 
Physical Church Turing Thesis (PCTT) is 
correct \cite{Kleene:1952},\cite{Deutsch:1985},\cite{Geroch:1986iu}, 
and the universe can 
be simulated to arbitrary precision
%\footnote{The ``arbitrary precision" qualifier 
%may be dropped if a version of digital physics  
%(such as the Church Turing Deutsch Principle) is correct. 
%This appears to be the case 
%if holography is correct (so that a finite sized volume
%contains the finite amount of information 
%tspecifiable on its boundary) .} 
inside a sufficiently powerful computer
(e.g. in a universal Turing machine $\mathcal{T}$).
%\footnote{With a Turing machine $\mathcal{T}$ specified by the 
%symbols $\Gamma$ it manipulates (we will generally use $\Gamma = \{0,1\}$ 
%which was shown sufficient by Shannon), the internal 
%states of the machine $Q$, and the transfer function $\delta$ which 
%describes what to do next given a certain symbol and state 
%(with some special considerations for the initial data and 
%state, and the halting state).  
%We will generally consider a simple universal 
%Turing machine $\mathcal{T}$ (22 internal states 
%is sufficient for $\Gamma = \{0,1\}$), so that 
%the majority of the ``program" is encoded in the 
%binary string $s_i$ that $\mathcal{T}$ begins with.
%In general, $s_i$ can include both a program 
%description (say the algorithm that 
%generates the Mandelbrot set: $z_{n+1}=
%z_{n}^2 + c$), and the initial data 
%that the program starts with (in this case, a set of 
%points $z$ from the complex plane).},
%\footnote{Note that we are not worried about computational 
%complexity here -- if the simulation is 3-EXPTIME, 
%so be it (not that we are proposing that this is the case -- presumably it 
%s something like BQP for a quantum universe). }).
This confidence is based both on the wide range 
of individual scientific successes to date, 
and the impressive way in which they 
overlap to form a unified whole.
By calculating of all of the 
interactions among the constituent particles
in a simulation  $\mathcal{S}$  %$SIM$
of a large region of the universe, 
one could thereby reproduce
stellar evolution and the formation of planets, 
the emergence of replicating molecular 
structures, and after a long period 
of Darwinian evolution the arrival 
of complex life forms which can observe 
and understand their environment.

If all of the details in 
the history of events inside a particular 
simulation $\mathcal{S}_i$ 
happen to match the evolution within our 
Hubble volume, 
then the observers within that simulation 
would be indistinguishable from us:
the patterns generated by their simulated 
neurons would precisely match the patterns 
formed in our brains.
At the same time, and at the deep level of the 
hardware generating the simulation, those 
observers are just another type of information: 
complex sequences of $1$s and $0$s that evolve
according to the implemented algorithm.
Everything else in that simulation, or in any other 
running program, is also just information -- different 
sequences of binary data.

Describing everything that exists in terms vast numbers 
of logical operations is admittedly a strident form of reductionism,
and is of little practical use in most day to day 
activities, scientific or otherwise.  
One primary purpose of this gedankenexperiment
is to help shake off the natural 
tendency to take existing as an observer 
for granted: those observers in the giant 
simulation are completely equivalent to 
the information encoded in  
sequences of $1$s and $0$s 
(as is everything else that exists).  
What is the special 
property of the observer-sequences so that one 
should exist as one of them, rather than some 
other form of information?

Before proposing an answer, let us first 
examine the nature of information in 
a bit more detail.
Note that the precise form of the 
various sequences of $1$s and $0$s is 
not necessarily critical.  
For instance, a Turing machine $\mathcal{T}_1$ 
running the simulation $\mathcal{S}_i$ can 
itself be emulated by a second universal 
Turing machine $\mathcal{T}_2$.   
The various objects inside the simulation $\mathcal{S}_i$ 
(say, comets and seagulls and 
picnics and so forth) will be encoded in 
different binary sequences in  
$\mathcal{T}_1$ and $\mathcal{T}_2$, but the 
real information content of these objects does not change.

There is thus a degree of redundancy possible 
when describing objects in a 
computational framework 
(or any other framework).
Consider the set $ \mathcal{A} $ 
given by the union of the output of all possible programs.
$\mathcal{A}$ can be generated by acting 
on all finite length binary strings $s_i$ with a 
universal Turing machine $\mathcal{T}$:
\be
\mathcal{A} = \bigcup \mathcal{T}(s_i) .
\ee  
There will then be an equivalence class $[\mathcal{S}_i]$ 
in $ \mathcal{A} $ for all of the programs 
that encode the same history of events described by $\mathcal{S}_i$:
\be
[\mathcal{S}_i] = \{\mathcal{S}_j \in \mathcal{A} | \mathcal{S}_i \sim \mathcal{S}_j \}.
\ee
More broadly, the elements of $ \mathcal{A} $ can be partitioned 
into the ``information structure" blocks $[x_i]$:
\be
\mathcal{A} ~/ \sim_{en} = \{[x_i] | x_i \in \mathcal{A} \}
\ee
where we use 
the special equivalence 
relation $\sim_{en}$ which has the abstract 
definition: ``encodes the same information".
For instance, in addition to detecting and grouping 
together all the programs that give rise to the simulation 
$\mathcal{S}_i$, this equivalence relation will also group together the 
data streams of the programs that give a proof 
of the Pythagorean theorem, 
or describe the life cycle of a species of fern, and so forth.
%(or, say, those that give instructions for baking a cake -- 
%some peculiar programs in $\mathcal{A}$ will 
%do both of these, and thus will need to be appropriately split).
We will later discuss the feasibility of explicitly 
constructing such a general equivalence relation.
%\footnote{In short, we argue that just as any particular 
%program $\mathcal{T}(s_a)$ will either halt or not 
%(but there is no finite length algorithm that determine 
%which is the case in general), that the 
%partition $\mathcal{A} / \sim_{en}$ does exist, 
%but the partition can not be completely 
%described by any finite length algorithm.}.

%for the content of the simulation 
%within the set possible possible programs.
%set of binary sequences $\{ Seq \}$.

Alternatively, a particular binary representation $x_{a1}$ for some 
information structure $x_a$ encoded on $\mathcal{T}_1$ 
amounts effectively to   
a gauge choice for the description of that structure 
-- similar to choosing 
a coordinate system $g_{ab}$ to map a 
spacetime manifold $\mathcal{M}$.
Furthermore, we will argue, in an analogy 
with the Faddeev Popov procedure 
(\cite{Faddeev:1967fc},\cite{Zee:2003mt}), 
that it is useful to get rid of the redundant 
gauge descriptions when counting 
among various forms of information.
We will therefore borrow the term 
``gauge" (in a slight abuse of the term) 
to describe the various representations 
that are possible for any information structure. 
%\footnote{Note however that all coordinate 
%gauges for a manifold $\mathcal{M}$ are simply 
%related by $\bar{g}_{ab} = {g}_{cd}\frac{\partial x^c}{\partial \bar{x}^a}
%\frac{\partial x^d}{\partial \bar{x}^b}$, while the gauge 
%transformations for describing a piece 
%information can be arbitrarily complex.}.
We are primarily interested in the translationally 
invariant information $x_a$, with various gauge choices 
$G(x_a)$ being more or less helpful in describing it. 

%To make a loose analogy with GR, 
%the invariant information is like a manifold $\mathcal{M}$,
%which can be described by various different coordinate 
%systems $g_{ab}$.
%We are primarily interested in this translationally 
%invariant information.  

%\footnote{The precise sequence $1$s and $0$s
%is not too important, as the same information 
%can be encoded on a different 
%architecture as a different sequence of 
%$1$s and $0$s.  We are interested in 
%the translation invariant information.}.

%It isn't necessary to actually build a giant 
%computer and run $\mathcal{S}$, 
%as the gedankenexperiment is sufficient to 
%demonstrate that any observer 
%can be completely represented as a form of information.
%Furthermore we will make the simplifying assumption 
%that everything else that exists 
%is also isomorphic to some type of information.

%Before proposing an answer, we will spend a bit more  
%time considering the range of information types.

We have assumed that the PCTT is correct, and thus everything  
that exists can be represented as various types 
of information $x_i$ (for convenience we will drop the brackets from $[x_i]$).
We will name the union of 
all these structures the universe of all information $\mathcal{U}$:
\be
\mathcal{U} = \bigcup x_i = \mathcal{A}~/ \sim_{en} \label{u_def}.
\ee
$\mathcal{U}$ resembles the von Neumann universe $V$ 
of set theory (which is a proper class \cite{Jech:2006}), 
or Tegmark's Level 4 Multiverse \cite{Tegmark:2003sr},\cite{Tegmark:2007ud}.
We adopt a Platonic viewpoint and assume that the 
entirety of $\mathcal{U}$ actually exists.
%\footnote{If not, what is the special principle promotes a subset 
%of $\mathcal{U}$
%(which would include at least the universe we find ourselves in) 
%from merely being possible to actually existing?}.
$\mathcal{U}$ thus contains mathematical structures like 
$x_{3p} = \{3 \in P\}$ (where $P$ are the primes), 
$x_{eul} = \{ e^{i\pi} + 1 = 0\}$, 
the Pythagorean theorem $x_{pt}$, the 
Mandelbrot set $x_{ms}$, the wave equation
$x_{we}$ and so on.

$\mathcal{U}$ also contains very complex structures 
like our physical universe, 
which we will denote by $\Psi$ (i.e. $\Psi \subset \mathcal{U}$).
Through objects like $\Psi$ the universal set will also contain all of the 
complex emergent phenomena that can arise 
through the interactions of many particles, such as 
volcanos $x_{vol}$, grasshoppers $x_{gh}$,
constitutional monarchies $x_{cm}$, 
and sensory qualia $x_{sq}$.
These emergent structures are 
somewhat statistical in nature, 
as defining them will involve some degree of 
coarse graining over microscopic 
degrees of freedom.

Observers are included among these complex structures, 
and we will grant them the special name $y_j$ 
(although they are also  
another variety of information structure $x_i$).
For instance a young child $y_{c1}$ may know about  
$x_{3p}$ and $x_{gh}$: 
$x_{3p}, x_{gh} \in y_{c1}$, while having not yet 
learned about $x_{eul}$ or $x_{cm}$.
This is the key feature of the observers that we will utilize:  
the $y_j$ are entities that can absorb various 
$x_i$ from different regions of $\mathcal{U}$.

\begin{figure}
\includegraphics[width=3.in]{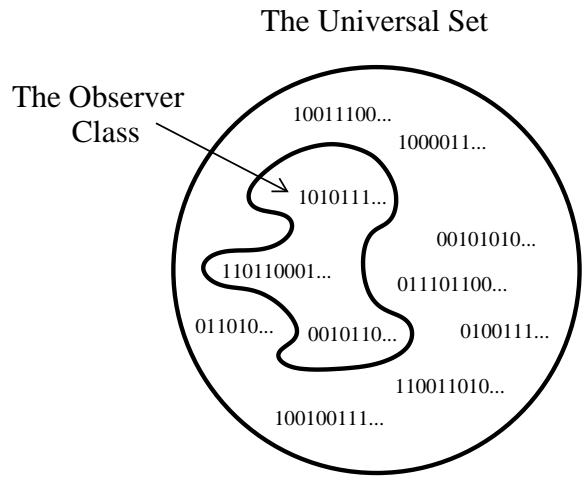}
\caption{The Observer class as a subset of the Universal Set 
(represented in some computational gauge).}
\label{ochfig1}
\end{figure}

We can now rephrase our central question in its 
final form.  Consider the Universal Set  $\mathcal{U}$: 
the set of all forms of information.
All of the observers $y_j$ will collectively form a particular subset 
of the Universal Set: the Observer Class $\mathcal{O}$:
\be
\mathcal{O} = \bigcup y_j \label{o_def}.
\ee
It is a given that to exist at all entails existing as some 
form of information -- to be some element 
within $\mathcal{U}$.  
Why is it that one exists as an element 
of the Observer Class $\mathcal{O}$ in particular?
This question is shown pictorially in Fig. \ref{ochfig1}, where the different
forms of information are symbolically represented 
as various strings of $1$s and $0$s 
(i.e. some computational gauge has been chosen).

The Observer Class Hypothesis (OCH) proposes a statistical answer: 
observers form by far the largest subset of $\mathcal{U}$:
\be
|\mathcal{O}| \gg |\mathcal{U}-\mathcal{O}| \label{ob_dom}
\ee 
and thus a element chosen randomly from the universal set 
is overwhelmingly likely be an observer 
(i.e. Fig. \ref{ochfig1} is not drawn to scale).
%This is illustrated in Fig. \ref{ochfig2}, where the observer 
%class has been qualitatively rescaled.
Just as typical observers dominate 
the counting among all observers, observers themselves 
dominate the counting among all forms of information.

%\begin{figure}
%%\includegraphics[width=3.5in]{OCH_dia2.pdf}
%%\includegraphics[width=3.5in]{OCH_dia2.eps}
%\includegraphics[width=2.2in]{teps4.eps}
%%\includegraphics[width=3.in]{teps5.eps}
%\caption{The Observer class $\mathcal{O}$ 
%is by far the largest subset of the Universal Set $\mathcal{U}$.
%An element drawn randomly from $\mathcal{U}$ will most 
%likely be from $\mathcal{O}$.}
%\label{ochfig2}
%\end{figure}

The observers form the most numerous class of information 
due to their primary trait: they can observe other 
forms of information, and thus incorporate them as subsets of themselves.
For $N$ different pieces of information $x_i$ in $\mathcal{U}$, 
there will generally be $2^N$ different possible observers $y_j$ 
who absorb
anywhere from $0$ to all $N$ of the $x_i$.
The observers thus collectively resemble the power set $\mathcal{P}(\mathcal{U})$ 
of the set of all information, and indeed we will later 
propose specifically that they effectively form the nontrivial power set 
$\hat{\mathcal{P}}(\mathcal{U})$:
\be
\mathcal{O} \sim \hat{\mathcal{P}}(\mathcal{U}) .
\ee
Note also that observers can observe other observers 
(and so on recursively) and so 
for example $\mathcal{P}(\mathcal{P}(\mathcal{U})) \subset \mathcal{O}$. 
We therefore refer to $\mathcal{O}$ as a class (i.e. a proper class) rather than a set.
In summation it is the combinatorial potential of the observers which 
leads them to vastly outnumber all other types of information, 
and thus provide the general form that existence invariably takes.

%%%%%%%%%%%%%%%%%%%%%%%%%%%%%%%%%%%%%%%%%%%%%
%%%%%%%%%%%%%%%%%%%%%%%%%%%%%%%%%%%%%%%%%%%%%
%              G  A  U  G  E  S
%%%%%%%%%%%%%%%%%%%%%%%%%%%%%%%%%%%%%%%%%%%%%
%%%%%%%%%%%%%%%%%%%%%%%%%%%%%%%%%%%%%%%%%%%%%
\section{Gauges and Counting}

%The observer class hypothesis is a very broad proposal, and 
%several problems appear to quickly crop up.
The core idea of the Observer Class Hypothesis 
is that if one counts over all forms 
of information they will find that observers 
greatly outnumber all other structures.
However, the counting process seems to be problematic 
at first for a number of reasons.
For example, there is the issue of over-counting pieces of 
information that have merely been 
re-expressed in a different form.
It would also appear that completely random 
structures in fact dominate the 
counting instead of observers.
These issues are solvable,
which can be demonstrated through 
an examination of   
gauge choice (i.e. the selection of 
some formalism to represent different 
types of information).
The concept of gauge invariance then 
leads to a more precise definition 
of information, and influences 
the consideration of the number of 
different forms of information that exist.

We first examine the gauge 
redundancy that exists in the representation 
of all forms of information.
Consider as an example the small element
 $x_{a1}$ inside of $\mathcal{U}$:
$x_{a1} = \{ 3 \in P \}$, where $P$ are the 
prime numbers.  This simple mathematical factoid 
$x_a$ can be re-expressed in an infinite number of 
different ways: $x_{a2} = \{ 1+2 \in P \}$, 
$x_{a3} = \{ \sqrt{9} \in P \}$, and so on 
(i.e. $x_{a1},x_{a2},x_{a3} \in G(x_a)$).
The proposal thus appears to be plagued by infinities, 
as even one simple mathematical statement 
would get counted an infinite number of times.

This over counting problem is not limited to 
the mathematical equivalence class.
To pick a literary example, one can take a famous quote from Hamlet, 
and arbitrarily translate it into the ``1-q" language:
``qto qbe qor qnot qto qbe", or into the 
``2-q" language and so forth.
In general this form of vacuous 
translational redundancy exists for all 
information structures.  
This can be seen in the example we noted before,
where a simulation $\mathcal{S}_i$ 
running on the Turing machine $\mathcal{T}_1$ 
can be emulated by a second universal 
Turing machine $\mathcal{T}_2$.  
The history of events within $\mathcal{S}_i$ won't change, 
but the sequence of $1$s and $0$s describing it  
in $\mathcal{T}_2$ does.

 It also appears at first that the universe 
 $\mathcal{U}$ is in fact dominated by 
 random noise.
 If we continue with the computational 
 framework and consider the elements of 
 $\mathcal{U}$ to be binary strings,  
 then there are $2^N$ different 
 strings $s_i$ that are $N$ bits long.
 The vast majority of these $2^N$
 strings are completely random, and do not encode 
 any real gauge-invariant information.
 This is reflected, for instance, in the fact that 
 the Kolmogorov complexity $K(s_i)$ 
 \cite{Solomonoff:1960},\cite{Kolmogorov:1965}  
 of an average string $s_i$ is comparable to 
 the length of $s_i$: 
 \be
 K(s_i) \sim |s_i|
 \ee
 %\footnote{As another example, consider a very large
 %sequence of data $X_M$, $M$ bits long, 
 %produced by our universe simulation
 %$\mathcal{S}$.  Let $M = N \cdot 2^N$ 
 %for some large value of $N$, so that 
 %$X_M$ in principle could contain 
 %all permutations of strings $N$ bits 
 %long.  For instance a 
 %1-petaflop machine 
 %could cycle through all 70 bit 
 %sequences over the coarse of a year.  
 %In practice however only a tiny 
 %fraction of the $2^N$ different 
 %sequences would occur in $X_M$ 
 %(unless a very inefficient coding gauge is used 
 %which artificially forces all of the permutations 
 %to occur).}.
 
 We will denote a true random string 
 (i.e. it has no internal patterns or real 
 structure) by $r_{i}$.
 This is distinct from the Kolmogorov definition 
 of randomness, which equates being 
 incompressible with being random.
 Most of the incompressible 
 $s_i$ strings are indeed patternless $r_i$ 
 strings, but a nontrivial fraction will instead  
 be the most compact representations of 
 real information structures 
 (e.g. consider the most compact binary description 
 of the Einstein equations $x_{EE}$ versus a random string 
 $r_i$ of the same length).
 
 We propose that both of these problems -- over-counting due 
 to gauge redundancies and the apparent 
 numerical dominance of the random 
 noise structures -- can be solved by choosing 
 a gauge.  Consider a similar problem 
 that arises in quantum field theory due to gauge 
 fields.  
 A straightforward evaluation of the path integral $Z$:
 \be
 Z = \int DA e^{\frac{i}{\hbar}\int \mathcal{L}(A,\partial A) d^4 x} \label{path_def}
 \ee 
  for a gauge 
 field $A_{\mu}$ gives an infinite result since the Lagrangian $\mathcal{L}$
 is invariant under gauge transformations 
 $A_{\mu} \rightarrow A'_{\mu} = A_{\mu} + \partial_{\mu} \phi$.
 Faddeev and Popov give us a solution: separate out the 
 redundant gauge portion of the Lagrangian 
 and integrate it separately as an irrelevant 
 infinite constant.
 Colloquially speaking, we get the correct 
 answer if we only consider the 
 physical variations of $A_{\mu}$, and not the 
 redundant gauge variations $G(A_{\mu})$.
 
 We posit that the same thing is true when 
 counting over all forms of information: 
 we need to 
 count over only the real, distinct information 
 structures (such as $x_{a1} = \{3 \in P \} $), 
 and ignore the redundant translations 
 (such as $x_{a2} = \{2+1 \in P \} $ and so forth)
 which would lead to an infinite over-counting.
 %\footnote{Actually it is not crucial for our purposes 
 %to have precisely one version of each 
 %information structure -- perhaps it would useful to also include 
 %$x_{a3} = \{\sqrt{9} \in P \} $.  We just want to avoid 
 %counting over things like 
 %$x_{a2270918} = \{ (517-280)/3-3^4+5 \in P \} $.}.
 The particular representation formalism chosen to describe the 
 different types of information is not important here, 
 as only its ability to remove the extraneous redefinitions is needed.
 
 We also postulate that the selection of a 
 representation gauge removes all of the random 
 noise structures $r_i$ which appear at first to dominate the 
 counting in $\mathcal{U}$.  In short, the random structures are 
 all gauge translations of nothing at all. 
 Consider for instance the quote from Hamlet, 
 which we can arbitrarily translate
 into the ``318-q" language, which makes the quote 
 much longer but doesn't add any real information to it.
 In the same fashion, a very long (and internally structureless) 
 random sequence $r_i$ is nothing more than the $r_i$ translation 
 of the null set $\O$:
 \be
 r_i \in G(\O)
 \ee
 The intuitive feeling that random structures don't contain any real, 
 nontrivial information is indeed correct, 
 and a manifestation of this is that they do not 
 carry any gauge-invariant information 
 (see also \cite{Chaitin:1990},\cite{Tao:2005}).
 
In contrast, non-trivial forms of information contain internal patterns 
and structures that map unambiguously from
representation to representation:
a child will learn that a mouse is a type of mammal 
in an English, French, or Chinese biology book.
%\footnote{There will be many be many subtleties 
%involved in the translation from one 
%real language to another (and these nuances 
%are themselves real forms of information), 
%unlike in the case of our ``1-q" language.}.
In comparison, one is free to translate a book-length 
random sequence $r_i$ into any other 
random sequence they wish, as it does not 
have any internal structures that must be 
preserved. 
We thus conclude that the true information content of the 
book $I(s_{bio})$ is much larger than 
the information content of a random string of the same length 
: $I(s_{bio}) \gg I(r_i) \sim 0$.
This is opposed to the result one gets from the 
Kolmogorov measure, where 
$I_K(s_{bio}) < I_K(r_i)$ since the biology book will be  
compressible to a degree, while the
random sequence $r_i$ is not.
Focusing on the gauge invariant information is therefore 
equal to selecting objects that have nontrivial 
``effective complexity" \cite{Gellmann:1996} or a 
large amount of ``logical depth" \cite{Bennett:1988},\cite{Ay:2008}.
As the random and redundant elements are removed 
by the selection of a gauge, we can then use the 
Kolmogorov measure to describe the complexity 
of the various $x_i$ (e.g. $I_K(s_{bio})$ for the biology book).
There will then be many more $x_i$ with large values for 
$I_K$ than those with a small amount of complexity, 
and thus a randomly 
chosen information structure will likely be very complex.

%That said, the Kolmogorov measure $I_K(s_{bio})$ 
%is a good estimate of the total quantity of information 
%in the book $I(s_{bio}) \sim I_K(s_{bio})$.
%Thus in general we can use the 
%Kolmogorov measure to estimate the size 
%of various $x_i$ once we have first removed 
%all of the redundant and random structures.
%To use different terminology, 
%the $x_i$ are thus the sparse subset of the $s_i$
%that have nontrivial ``effective complexity" \cite{Gellmann:1996} or a 
%large amount of ``logical depth" \cite{Bennett:1988},\cite{Ay:2008}.
%In general there will be many more 
%$x_i$ with a large Kolmogorov complexity than a 
%small one (via the pigeonhole principle).

We have described the real, non-trivial 
information structures 
as having gauge invariant internal patterns, 
but have not given an explicit 
characterization of this property.
To this end one imagines a finite length algorithm $F$ that 
 can search through all 
$2^{N}$ of the $N$ bit strings 
$s_i$ (for any $N$) and extract and 
organize 
all of the gauge invariant information 
structures $x_i$, with the remainder 
being the gauge translations of those 
structures $G(x_i)$, or the random 
strings $r_i = G(\O)$.
For instance $F$ would be able to examine the 
strings $s_p = \{ 3 \in P \}$
and $s_q = \{ 4 \in P \}$, and select 
$s_p$ to be filed among 
the $x_i$, while discarding $s_q$,
or determine that the
effective complexity of a frog  
is much greater than a 
glass of water of the same mass.

%This would give an alternative
%form for $\mathcal{U}$:
%\be
%\mathcal{U} = \frac{\bigcup s_i}{G}.
%\ee 

As a concrete example, consider the 
simple world of binary addition $\psi_{BA}$, 
which is composed of strings of 
4 symbols: 0,1,+,=.  
In general it is possible to have correct strings:
``1+1=10", incorrect strings: ``1+0=11", 
and ill-formed strings: ``+1+==".
For this toy model one can construct an algorithm 
$F_{BA}$ that will search through 
all $4^N$ of the $N$ symbol strings 
and extract the fraction which are correct.
A numerical implementation of $F_{BA}$ shows 
that the number of correct strings 
scales approximately as $\sim 0.17 \times 2.53^N$.
This is the type of behavior that we expect in the general case:  
the number of real structures will grow 
exponentially (via combinatorics) 
as the string length $N$ increases, 
but at a slower rate than the total number 
of strings. 
The $x_i$ will thus form a sparse 
subset of the strings $s_i$, with a density 
trending towards zero as $N \rightarrow \infty$.
On a side note, we do not conclude that 
there is an infinite amount of real information 
in $\psi_{BA}$, even though there are 
correct strings of unbounded length.
Rather, the amount of 
nontrivial information in $\psi_{BA}$ is comparable 
to the length of the algorithm $F_{BA}$: 
$I(\psi_{BA}) \sim |F_{BA}|$.

One can construct these types of algorithms for various 
simple problems, but it is not possible to construct 
a finite length algorithm $F$ 
for the general case, as demonstrated 
by G\"{o}del's Incompleteness Theorems \cite{Nagel:2008}, 
and Church and Turing's negative answer to 
Hilbert's Entscheidungsproblem \cite{Church:1936},\cite{Turing:1937}.
This itself is not a problem, but 
rather a reflection of the natural structure of $\mathcal{U}$.
Just as random strings $r_i$ 
can't be compressed, 
the union of real information structures (\ref{u_def})
%\be
%\mathcal{U} = \bigcup x_a
%\ee  
can not be completely encapsulated
by any finite length description.
Ignoring categorization, there is not even 
a universal trait that all 
of the individual $x_i$ carry which 
a ``litmus test" type $F$ could test for:
no single equation in Wiles' papers 
proves Fermat's Last Theorem, no single 
chemical bond in a chlorophyll molecule 
shows that photosynthesis is a useful 
evolutionary adaptation.
One needs essentially a library (or the internet)
to give a zeroth order approximation 
of $F$ for merely the things we have 
learned so far in our corner of $\Psi$.
The best one can do for a concise 
measure of information content 
is thus something like the 
Kolmogorov measure, which 
works due to the pigeonhole 
principle, but is necessarily 
silent on the semantic content of the various 
binary strings.

Explicitly constructing a finite algorithm 
$F$ that sifts through all binary 
strings $s_i$ and extracts all of the $x_i$
which compose $\mathcal{U}$ is not 
possible, but in the end it isn't needed either.
%The sparse subset of the $x_i$ contains all of the 
%true information embedded among the $s_i$ strings, 
%with the $r_i$ being the remainder - 
%the chippings left over after carving 
%the $x_i$ from the marble block of $s_i$.
To make an analogy, 
$\Psi$ does not actually physically perform a  
Faddeev Popov type procedure to remove the 
redundant gauge degrees of freedom 
every time two gluons interact
(which alternative formulations like BRST show).
Alternatively, all of the various programs $\mathcal{T}(s_i)$ 
in $\mathcal{A}$ will either halt or not, there just isn't 
a finite length algorithm that can determine which is 
the case in general.   
Likewise, $\mathcal{U}$ simply has a natural, intrinsic structure, 
which the consideration of gauge invariance merely 
helps to highlight.
This natural structure is clearest in the case 
of the mathematical $x_i$: 
it really is the case that 3 is a prime number 
and 4 is not (see also \cite{Gowers:2000},\cite{Tao:2007}).
Just because we consider the union of all 
$x_i$, it is emphatically not the case that 
``anything goes".

The existence of this natural structure 
also carries over to the 
complex emergent phenomena that can 
arise in physical universes like $\Psi$.
For instance, a salamander ($x_{sal}$) is a nontrivial  
information structure 
which occurs in our universe.
One consequence of being a real $x_i$ 
is that salamanders exist much more frequently  
in our Hubble volume than a direct counting 
of all the permutations $\Omega$ 
of the same number of atoms would suggest.
In fact, we will later use a high frequency of 
occurrence $f(x_i)$ 
as an indicator that an emergent object $x_i$ has
nontrivial structure.
%(as compared to a background rate $f_{perm}(x_i) \sim N(x_i(m))/N(m)$
%given by the number of permutations of the same 
%amount of mass)

%In fact, we will use the frequency $f(x_i)$ at which 
%a object occurs 
%as an indicator  
%of the non-triviality of that object
%(where the object is defined with an appropriate level of coarse graining) .
%That is, if $f(x_i) \gg f_{perm}(x_i)$  
%(where $f_{perm}(x_i)$ is the 
%expected frequency based on a straight 
%forward counting of all permutations 
%of the same mass) then we can 
%say that $x_i$ has non-trivial structure.
%We note that $f(x_i)$ is itself a function of the 
%internal structure of the object, which we can 
%not hope to describe in general.

The intricacy of these complex emergent 
phenomena can be seen if we 
further prod our example structure $x_{sal}$:  
that is, what do we mean precisely by a salamander?
A certain individual organism at one moment in time?
Do we coarse grain over the orientations of the  
water molecules in that individual's cells?
What about the particulars of the 
ongoing gene expression in those cells at that moment in time?
Do we instead consider all the individuals in a breeding population 
at a moment in time, or the general features 
of their shared genotype and phenotype
over a broader span?
Furthermore there has been a nearly continuous 
evolution (being discrete at the parent-child level) 
from the most basic replicating molecular structures 
early in the Earth's history to the various species 
that exist today (see, e.g. \cite{Dawkins:2005}) 
-- where does one make a cutoff?

Naturally, there is no single right answer, as different 
definitions and levels of detail illuminate different 
aspects of the information 
structure which we concisely named $x_{sal}$.
While very complex and inextricably interwoven 
with many other objects 
(and local to the branch of $\Psi$ we find 
ourselves in), 
$x_{sal}$ is also not arbitrary or random: it is an 
example of a natural structure occurring in $\mathcal{U}$, 
not unlike the mathematical structures $x_a = \{ 3 \in P\}$
or $x_b = \{ e^{i\pi} + 1 = 0 \}$.
Note, however, that in fact all possible arrangements 
of atoms will occur with at least some non-zero 
frequency inside an eternally 
inflating $\Psi$.
Just existing as a pattern somewhere within a physical universe like $\Psi$ 
is thus too imprecise of a measure for membership 
in $\mathcal{U}$.
Rather, we generally want the patterns that occur frequently in 
$\Psi$ to be promoted to the $x_i$ elements of $\mathcal{U}$ -- 
those patterns that minimize the Lagrangian in (\ref{path_def}), 
roughly speaking. 

In fact, there is a natural solution for 
selectively ranking the various complex patterns 
and phenomena that 
can arise in a universe like $\Psi$: do not 
consider the patterns in perfect isolation 
(like pure mathematical factoids), but rather as 
the emergent structures embedded in 
a physical universe that they are.
The frequency $f$ at which a particular 
pattern occurs in $\Psi$ 
(which is closely linked to the process that generates it)  
thus becomes a useful tool in 
measuring its information content.
For instance, all sorts of bizarre objects 
(say, interplanetary china teapots: $x_{ict}$) 
can be formed in an eternally 
inflating $\Psi$ 
through macroscopic quantum tunneling events.
As these objects are formed through 
abrupt random processes they contain 
little gauge-invariant information, 
and this in turn is linked to their 
exponentially small production rates 
(which also corresponds to having very little 
logical depth \cite{Bennett:1988}).
%but being exponentially unlikely these 
%types of randomly generated structures have 
%very low information content.
In short, they can be quickly and concisely described 
as very improbable 
random fluctuations, and thus for 
example: $I(x_{sal}) \gg I(x_{ict})$.

Note that the frequency at which a particular 
pattern $x_a$ occurs 
(which is a function of $x_a$'s internal structure) 
is only one component of 
its information content -- 
there is also the actual complexity $ \sim I_K(x_a)$ of the pattern 
(at an appropriately coarse-grained level).
The gauge invariant information contained in a pattern $x_a$
will thus roughly scale as the complexity of the pattern   
times the frequency of its occurrence:  
\be
I(x_a) \propto I_K(x_a) \times f(x_a).
\ee
$I_K(x_a)$ is the baseline 
Kolmogorov complexity of the pattern, and a large value for 
$f(x_a)$ then makes it likely that the value of 
$I_K(x_a)$ stems from nontrivial effective complexity 
rather than random noise 
(e.g. the compact description of the Einstein equations 
$x_{EE}$ will have a large value for $f$ as compared to a 
vanishing value for a random sequence $r_i$ of the 
same length).

%For example, perhaps it is the case that 
%there are the same number of 
%number 2 pencils $x_{no2}$ and 
%dolphins $x_{dol}$ on the planet (so that $f(x_{dol}) \sim f(x_{no2})$), 
%but the dolphins are much more complex 
%($I_K(x_{dol}) \gg I_K (x_{no2})$)
%and they will thus contain a greater amount of 
%gauge invariant information: $I(x_{dol}) \gg I(x_{no2})$.

In fact, it would generally suffice for our purposes 
to define $f$ to be a step function, 
so that it equals $1$ for those patterns 
that have a greater than 50 percent chance 
of occurring within a Hubble volume 
over the 100 billion years of standard 
stellar evolution, and zero otherwise.
In this way $f$ does the work of $F$ 
and effectively extracts the sparse 
subset of nontrivial structures from among the 
exponentially large set of possibilities.
For instance, if we coarse grain biological 
patterns at the cellular level 
(with, say, 100 different 
cell types, each with an average mass of $10^{-9}$ grams), 
then there are $\sim 100^{10^{14}}$ 
different possible arrangements for 
a $100$ kilogram ``organism".
In a similar fashion to the simple binary addition example $\psi_{BA}$, 
the number of viable organisms 
(i.e. $f = 1$) 
will grow 
exponentially with increasing mass, 
but at a much slower rate than the total 
number of arrangements 
(i.e most permutations have $f=0$).

A step function profile for $f$ is admittedly 
artificial -- in reality $f$ will be a function of the 
internal structure of the various patterns
(and the environment in which they occur).
In general, there is a smooth 
continuum from exceptionally unlikely to 
commonplace phenomena, and thus there 
is no clean cutoff such that we could include 
the more common objects 
to be among the real information structures $x_i$ while 
discarding the rest as random $r_i$.
Indeed, evolution via natural selection, lacking goals 
or foresight, must sift through many different 
patterns (of varying degrees of randomness) 
in a process of iterative trial and 
error in order to generate nontrivial 
emergent structures such as $x_{sal}$.
Likewise, even with goals and foresight, iterative 
trial and error is an indispensable  
component of scientific discovery -- 
as observed for instance by Poincar\'e.
We are thus not calling for the elimination  
all aspects of randomness, but rather merely positing that 
the apparent preponderance of the completely random 
$r_i$ in $\mathcal{U}$
was always an illusion. 

By including the frequency at which 
various complex emergent phenomena 
occur as a useful indicator   
of their information content we essentially 
rediscover the measure problem.
Within the confines of our Hubble volume things 
are clear cut: spontaneous violations of the 
2nd law are very rare (say, a glass of water 
separating out into ice and steam), and thus have a 
very low information content.
In the expanse of an infinite universe however, 
even these very rare events can occur an infinite 
number of times, and the counting thus 
becomes trickier.
It would seem that with a countably infinite 
number of both the ``rare" and ``common" events 
that one is free to arrange and count them 
in any fashion and thus arrive at any relative frequency.
A common example of this is to ask what fraction 
of the integers are even.  A natural guess 
is $1/2$: $\{ 1,2,3,4,5,6,\dots \}$, but they can be 
rearranged to give any fraction between 
zero and one -- say $1/3$: $\{ 1,3,2,5,7,4,\dots \}$.

In response to this we ask a related question: what 
fraction of the integers are prime?
The same trick can be played here, with the primes 
and composites being rearranged to give 
any desired fraction.  
However, if we make the 
restriction of considering all integers less 
than $N$, and then letting $N$ 
go to infinity, 
then there is a nontrivial answer: 
the number of primes $\pi (N)$
less than or equal to $N$ is approximately equal to $N/\ln (N)$
(or, more closely, Li$(N) = \int_2^{N} dt/\ln(t)$).
If the Riemann hypothesis is correct we can even 
add the order of magnitude error estimate:
$\pi(N) = $ Li$(N) + O(\sqrt{N}\ln (N))$.
Just asking what fraction of the integers 
are prime is thus an incomplete question.
The question is uniquely finished by adding the 
constraint (or regularization) of considering 
all integers less than $N$ (and then letting $N \rightarrow \infty$).
Furthermore, the correctness of this particular question can then be seen 
by the abundance of natural structure (the Prime Number 
Theorem $x_{pnt}$, and the associated properties of the 
Riemann Zeta function) that it leads to. 

We assert that the same scheme 
holds for physical universes like 
$\Psi$: they can be infinite in extent 
and still contain nonrandom, nontrivial 
structure (which can be revealed by asking the right questions). 
Agricultural societies, for instance, will 
occur much more frequently 
in $\Psi$ type universes 
than teapots spontaneously 
formed in protoplanetary disks.
This is certainly the case  
within the confines of our Hubble volume, and the puzzle 
is to then find the correct regularization or 
measure that extends this general pattern  
of relative frequencies to 
arbitrarily larger volumes.
Note that it is not the case that 
the relative frequencies in distant regions will be 
identical to those we observe locally 
-- but rather it should emerge 
that our Hubble volume is typical 
among the subset of regions 
that support complex life.
This problem is a good bit more subtle 
than the case of regulating the integers, 
and we will not offer  
our own measure proposal here.
We will also refrain from endorsing 
other measure proposals, although we 
think there are interesting candidates
that may be on the right track.

Our primary goal is even broader 
in scope than recovering the 
general distribution of probabilities 
within a physical universe like $\Psi$.
We claim that it is not only the case 
that the $x_{sal}$ type objects greatly 
outnumber the random $x_{ict}$ objects in a $\Psi$
type universe (i.e. we assume that a correct 
regularization which confirms one's intuition is possible), 
but that the $x_{sal}$ 
outnumber the 
$x_{ict}$  
throughout the entirety of $\mathcal{U}$
(and therefore the nonrandom   
structures dominate the information content of  $\mathcal{U}$).

To give a possible counterexample to this 
proposal, there will be ``heat bath" type 
universes $\psi_{HB}$ within $\mathcal{U}$, 
such that all of their internal patterns are generated 
through random fluctuations.
In these $\psi_{HB}$ universes the 
$x_{sal}$ and $x_{ict}$ type structures will 
be produced at comparable rates.
Since no consistent patterns emerge, 
the interactions that occur within a spacetime 
volume of one of these $\psi_{HB}$ universes 
can effectively be compressed to the initial state and 
evolution equations that then evolve it.
$\Psi$ on the other hand 
starts with a very low initial entropy (instead of 
a maximal entropy like the $\psi_{HB}$), and will 
thus preferentially generate certain complex emergent 
patterns (the ``viable ones") as the entropy increases 
(and these emergent structures
are not described by the initial 
conditions and evolution equations, 
other than in some ``latent" sense).
The heat bath universes are thus effectively  
the analogue of the long random strings $r_i$: 
they appear very complex at first, but they carry 
little real information -- essentially only that 
carried in their evolution equations.
%However, just as a random binary sequence $r_i$ 
%initially appears to carry a lot of information 
%due to being incompressible, but in fact carries  
%no real, gauge-invariant information, 
%the characterization of a $\psi_{HB}$ type universe seems at first to require 
%very long description, but in fact 
%it can be succinctly and completely described 
%as random heat bath.
The information content $I(\Psi)$ within the $\Psi$ type 
universes is thus vastly greater 
than the random $\psi_{HB}$ type 
universes: $I(\Psi) \gg I(\psi_{HB})$.

In addition to the completely disjoint 
$\psi_{HB}$ type universes existing elsewhere within $\mathcal{U}$, 
it is possible that our own universe could evolve into 
an eternal heat bath phase.
This would be the case, for instance, if the current acceleration 
is due to a true cosmological constant $\Lambda$, 
so that we will eventually transition 
to an eternal de Sitter universe, 
with a horizon temperature of:
$T_{dS} = \sqrt{2G \rho_{\Lambda}/3\pi}$.
A straightforward counting of all of the 
observers $y_j$ in this scenario would seem to be 
dominated by the infinite number of Boltzmann Brain 
observers $y_{BB}$ in the cold de Sitter phase.
In turn this is sometimes presented as evidence that $\Lambda$ 
must not stable and will decay at some point in the future.
There are independent reasons to be skeptical 
of an eternal de Sitter phase \cite{Dyson:2002pf}, and we 
are not proposing that our universe has a true, 
immutable $\Lambda$.
However, we do not think the possibility 
of an infinite number of $y_{BB}$
rules a true $\Lambda$ out: the real 
information is contained in the initial, 
stellar evolution phase of $\Psi$.

Finally, one can also conjecture the existence of other types 
of physical universes (i.e. $\Psi^{-1}$) where 
the internal dynamics (during an entropy 
increasing phase) specifically leads to the bizarre objects 
like the interplanetary teapots consistently outnumbering 
the natural objects like salamanders.
If these universes were to exist, then upon counting over both 
the $\Psi$ and $\Psi^{-1}$ one could find that 
all possible patterns are produced at comparable rates.
Here we simply assert, given the inherent  
structure of $\mathcal{U}$ and the deep 
mathematics that provide the 
foundation of $\Psi$, that these $\Psi^{-1}$ type 
universes do not exist -- just as the statements  
$\{ e^{i \pi} + 1 = 4 \}$, or ``the Mandelbrot 
algorithm produces an isosceles triangle" do not 
correspond to $x_i$ in $\mathcal{U}$.
The nonrandom emergent patterns that arise in 
nontrivial mathematical structures like $\Psi$ are therefore also 
the primary content of $\mathcal{U}$ itself.

In summation, the universe of all information is full 
of nontrivial mathematical structures and complex 
emergent phenomena -- i.e. the sorts of things 
that observers are interested in absorbing.
Indeed, it is because to this ability to selectively extract 
information from other sources that observers 
themselves form the vast majority of $\mathcal{U}$.

%Just as any particular program 
%will either halt or not, and there is merely 
%no finite-length algorithm that can determine 
%which happens for all cases, 
%we assert that the real, gauge-invariant information 
%structures $x_i$ are the content of $\mathcal{U}$, 
%and there merely is no finite length algorithm 
%that can categorize them all.

%%%%%%%%%%%%%%%%%%%%%%%%%%%%%%%%%%%%%%%%%%
%%%%%%%%%%%%%%%%%%%%%%%%%%%%%%%%%%%%%%%%%%
%  OBSERVERS
%%%%%%%%%%%%%%%%%%%%%%%%%%%%%%%%%%%%%%%%%%
%%%%%%%%%%%%%%%%%%%%%%%%%%%%%%%%%%%%%%%%%%
 \section{Observers}
 
 The Observer Class Hypothesis proposes that 
 the observers $y_j$ collectively 
 form the largest subset of the 
 universe of all information $\mathcal{U}$.
 In particular, the way in which the observers 
 selectively absorb $x_i$ from 
 various regions of $\mathcal{U}$
 (say, $x_a, x_b, x_d,... \in y_{\alpha}; 
 x_a, x_c, x_e,... \in y_{\beta};...$), results in the 
 observer class resembling the power 
 set $\mathcal{P}(\mathcal{U})$.
 We will refine this observation, 
 and propose specifically that observers 
 effectively form the nontrivial power set of 
 $\mathcal{U}$.  
 The ``non-triviality" restriction for  the elements of 
 $\mathcal{P}(\mathcal{U})$ 
 then naturally leads to physically 
 realized observers existing 
 within a superstructure like $\Psi$.
 We then consider the overall size of 
 $\mathcal{U}$, including the possibility 
 that it is infinite in extent.
 If $I(\mathcal{U}) = \aleph_0$, then universal observers 
 $\hat{y}_j$ may be able to absorb any 
 $x_i$ by continuously 
 upgrading their ability to process information.
 
 Consider first the straightforward power set $\mathcal{P}(\mathcal{X})$
 of some set of information $\mathcal{X} = \bigcup_{i=1}^N x_i \subset \mathcal{U}$.
 The amount of information in $\mathcal{X}$ can 
 be found by summing the information content of its members:
 \be
 I(\mathcal{X}) = \sum_{i=1}^N I_K(x_i)
 \ee
 (where one can use the Kolmogorov measure $I_K$ as the 
 alternative gauge versions $G(x_i)$ and random strings $r_i$ have already 
 been cut from $\mathcal{U}$).
 What then is the information content of $\mathcal{P}(\mathcal{X})$?
 Each element $x_a \in \mathcal{X}$ will occur in half of the 
 $2^N$ elements of $\mathcal{P}(\mathcal{X})$, so 
 a direct counting would give  $I_{dir}(\mathcal{P}(\mathcal{X})) = 2^{N-1} I(\mathcal{X})$.
 This is a very inefficient description however -- the entire 
 content of $I(\mathcal{P}(\mathcal{X}))$ is in fact contained 
 in the combination of the original set, and in the definition of the power set, so that 
 the true information content is: 
 $I(\mathcal{P}(\mathcal{X})) = I(\mathcal{X}) + I(\mathcal{P}) \sim I(\mathcal{X})$.
 
 However, it is also possible to combine basic 
 information structures and form a new, nontrivial structures 
 through their interconnections.
 For instance, various aspects of Riemannian geometry 
 ($x \in \mathcal{X}_{Riem}$)
 were used in the derivation of the Einstein equations of 
 general relativity, 
 or in Penrose and Hawking's singularity theorems, 
 and in Perelman's proof of the Poincar\'e 
 conjecture.  
 Different combinations of proteins 
 ($x \in \mathcal{X}_{prot}$) allow 
 assorted bacteria to live in environments ranging 
 from the root systems of fig trees, to the shores 
 of alpine lakes or in  
 thermal vents on the sea floor.
 Alternatively, various arrangements of electrical 
 components ($x \in \mathcal{X}_{elec}$) 
 allow for the creation 
 of radios, radars and integrated 
 circuits. 
 %All of the objects formed from the constructive 
 %combinations of simpler elements of some 
 %$\mathcal{X}$ 
 %(where repetition of the simple elements is often used) 
 %are effectively 
 %the elements of the nontrivial power set of $\mathcal{X}$.

 We therefore define the nontrivial power set 
 $\hat{\mathcal{P}}(\mathcal{X})$, 
 to be the sparse subset of 
 combinations from $\mathcal{X}$ that lead to new structures in $\mathcal{U}$:
 \be
 \hat{\mathcal{P}}(\mathcal{X}) = \mathcal{P}(\mathcal{X}) \bigcap \mathcal{U}.
 \ee
 There are a couple caveats in this definition. 
 In general the repetition of simple elements in $\mathcal{X}$ is 
 allowed (e.g. many transistors are needed for a computer chip).
 Additionally the elements of $\mathcal{X}$ may need a system 
 of tags so that the interconnections within the 
 new structures can be explicitly identified 
 (say, the locations of the circuit elements on 
 the computer chip).
 The straightforward power set of some set $\mathcal{X}$ 
 (of $N$ elements) does not add any new information, but the selection 
 of the nontrivial combinations does:
 \be
 I(\mathcal{X}) \sim I(\mathcal{P}(\mathcal{X})) \ll I(\hat{\mathcal{P}}(\mathcal{X})).
 \ee
 The elements of $\hat{\mathcal{P}}(\mathcal{X}) - \mathcal{X}$ are 
 new $x_i$ elements of $\mathcal{U}$ 
 (which observers can also absorb).

 Consider then the proposed observer $y_{r1}$ 
 (i.e. a direct element of $\mathcal{P}(\mathcal{U})$):
  $y_{r1} = \{ x_{tang}, x_3, x_{nept} \}$, 
 where $x_{tang}$ is a tangerine, $x_{3}$ is the 
 number 3, and $x_{nept}$ is the planet Neptune.
 This random collection of various information structures 
 from $\mathcal{U}$ is clearly 
 not an observer, or any other from of nontrivial information:  
 $y_{r1}$ is redundant to its three elements, and would thus 
 be cut by the selection of a gauge.
 This is the sense in which most of the direct elements of the 
 power set of $\mathcal{U}$ do not add any new real information.
 
 However, one could have a real observer $y_{\alpha}$
 whose main interests happened to include types of fruit, the integers, and 
 the planets of the solar system and so forth.
 The 3 elements of $y_{r1}$ exist as a simple list, 
 with no overarching structure actually uniting them.
 A physically realized computer, with some finite 
 amount of memory and a capacity to receive 
 input, resolves this by providing a 
 unified architecture for the nontrivial 
 embedding of various forms of information.
 A physical computer thus provides the glue to combine, say,   
 $x_{tang}$, $x_{3}$, and $x_{nept}$ and 
 form a new nontrivial structure in $\mathcal{U}$.
 
 It is possible to also consider the existence 
 of ``randomly organized computers" 
 which indiscriminately embed arbitrary 
 elements of $\mathcal{U}$ -- these 
 too would conform to no real $x_i$. 
 This leads to the specification of ``physically realized" 
 computers, as the restrictions that 
 arise from existing within a mathematical 
 structure like $\Psi$ results in  
 computers that process information in 
 nontrivial ways.
 Furthermore, a structure like $\Psi$ allows for 
 these physical computers to spontaneously 
 arise as it evolves forward from an initial state of 
 low entropy.
 Namely it is possible for replicating 
 molecular structures to emerge, and  
 Darwinian evolution can then drive to them 
 to higher levels of complexity as they 
 compete for limited resources.
 A fundamental type of evolutionary 
 adaptation then becomes possible: 
 the ability to extract pertinent information 
 from one's environment so that it can 
 be acted upon to one's advantage.
 The requirement that one extracts useful 
 information  
 is thus one of the key constraints that 
 has guided the evolution of the 
 sensory organs and nervous systems 
 of the species in the animal kingdom.
 
 This evolutionary process has reached its current 
 apogee with our species,
 as our brains are capable of extracting information 
 not just from our immediate surroundings, but also 
 from more abstract sources 
 such as $\mathcal{X}_{Riem}$, $\mathcal{X}_{prot}$, 
 or $\mathcal{X}_{elec}$.
 %In addition to observing and reacting to the transient phenomena in 
 %our immediate surroundings, we can investigate the 
 %causes and consistent patterns behind them.
 %We can observe the details of our immediate surroundings
 %like the other animals, but also actively investigate the 
 %assorted causes and patterns behind the various transient 
 %phenomena.   
 We can  
 absorb the thoughts of others on wide ranging subjects, 
 from Socrates' ethics, to 
 Newton's System of the World.
 %and Everett's Relative State Formulation.
 Observers can also create structure, from 
 individual works like van Gogh's 
 paintings, to collective entities like the 
 global financial system.
 The combinatorics that arise from the 
 expansive scope of sources that Homo Sapiens 
 can extract information from thus explains why 
 we are currently the typical observers. 
 %\footnote{In other words, one 
 %counts over the number of possible mental states, 
 %not just brains -- otherwise something like an ant would 
 %be the typical observer.}.
 
 We have hypothesized that observers comprise the largest class 
 of information, 
 but the implications that follow from this depend on 
 the size of $\mathcal{U}$ itself.
 In general there are three possibilities for the extent of 
 the universe of all information: 
 \begin{itemize}
 \item $\mathcal{U}$ is finite:  $I(\mathcal{U}) < N$ (for some finite $N$), 
 \item $\mathcal{U}$ is countably infinite: $I(\mathcal{U}) = \aleph_0$, 
 \item $\mathcal{U}$ is uncountably infinite: $I(\mathcal{U}) > \aleph_0$.
 \end{itemize}
 Roughly speaking, $\mathcal{U}$ could be 
 ``small", ``big", or ``very big".
 It could be argued that the first case is 
 the null assumption, but we will also 
 seriously examine the possibility that $I(\mathcal{U}) = \aleph_0$
 (which follows if there is no finite upper bound to the complexity 
 of real information structures).
 However, we will not inspect in depth the $I(\mathcal{U}) > \aleph_0$ 
 case (which would necessitate $x_i$ that are irreducibly infinitely 
 complex).  
 It is conceivable that the third case could be meaningful 
 if hypercomputers of some form could be built,  
 %\footnote{For instance, a machine that could 
 %perform each subsequent step in half the time 
 %of the previous one.}, 
 but we find this scenario unlikely.
 We note in passing that if was the case that 
 $I(\mathcal{U}) > \aleph_0$, then it may be possible 
 to demonstrate (\ref{ob_dom}) via cardinality arguments.
 %In any case $I(\mathcal{U}) = \aleph_0$ is a large enough 
 %structure to contemplate for the time being.
 %(while only briefly inspecting $I(\mathcal{U}) > \aleph_0$).
 
 First consider the case that $\mathcal{U}$ is finite.
 One begins with an infinite number of 
 programs producing data in $\mathcal{A}$, but the action 
 of removing both the redundant translations 
 and the random output significantly prunes 
 the binary tree.  
 It is possible that after this gauge cut only a finite number of 
 finitely complex structures $x_i$ remain, 
 so that the sum of all their information 
 content is also a finite number.
 Note that this is possible since we use compact representations 
 for the various $x_i$. 
 For instance, one could go through and list 
 all of the prime integers, thereby generating an infinite 
 number of true statements, but the existence of  
 an infinite number of primes is concisely expressed in 
 Euclid's proof.
 Likewise, a single uncomputable real requires 
 an infinitely long description, 
 %\footnote{If there are no infinitely complex real structures, then 
 %the uncomputable reals are essentially 
 %infinitely long random $r_i$.}, 
 but the general procedure 
 to generate reals can be concisely defined by Dedekind cuts 
 (therefore most real numbers are effectively infinitely long $r_i$ sequences).
 %Alternatively, there are many different types of infinite 
 %cardinalities proposed in set theory, 
 In this sense there may only be a finite number 
 of unique mathematical structures, 
 encapsulated in some $\mathcal{X}_{math}$, 
 so that no combinations from $\mathcal{P}(\mathcal{X}_{math})$ 
 lead to new, nontrivial structures.
 %\footnote{That is, we lean towards ontological 
 %maximalism, but it may be that there are only a finite 
 %number of finitely complex objects in $\mathcal{U}$.}. 
 
 %(and there are no other 
 %``atomic" structures not included in $\mathcal{X}_{math}$).
 
 Furthermore, if $I(\mathcal{U}) < N$, then our physical 
 universe $\Psi$ only contains a finite amount of 
 real information.  $\Psi$ could still be infinite in 
 spatial extent, perhaps due to a form of eternal inflation.
 However, in this scenario, the dynamics of $\Psi$ would only allow finitely complex 
 objects to arise in any one region. 
 For instance, the deepest dynamics of $\Psi$ may be completely described  
 by some finitely complex mathematical 
 structure (perhaps a completion of string theory).
 These fundamental dynamical laws may only admit a finite number of 
 degrees of freedom in any finite volume of spacetime 
 (in agreement with current Planck scale/Holography proposals 
 \cite{'tHooft:1993gx},\cite{Susskind:1994vu},\cite{Polchinski:2010hw}).
 The construction of arbitrarily complex structures would 
 then necessitate coherent communication across arbitrarily large scales.
 If it is the case that $I(\mathcal{U}) < N$, then 
 unbounded communication of this sort is not possible.
 Additionally, if $\mathcal{U}$ is finite, then necessarily there  
 are not arbitrarily more 
 complex alternative universes $\psi$ elsewhere in $\mathcal{U}$
 -- the complexity of the structures within our $\Psi$ 
 would be representative for the 
 finite number of other physical universes in $\mathcal{U}$.

 If the Observer Class Hypothesis is correct and 
 $\mathcal{U}$ is finite, then in principle a 
 direct counting would show 
 $|\mathcal{O}| \gg |\mathcal{U}-\mathcal{O}|$.
 In practical terms there would be physical
 observers $y_j$ with the memory capacity to absorb 
 a large number of the most complex $x_i$ in $\mathcal{U}$ -- 
 combinatorics would then ensure that the 
 observers dominate the counting.
 The number of $x_i$
 structures with a Kolmogorov complexity 
 of $N$ bits should then roughly look  
 like a rescaled and truncated Poisson distribution $f(N;\lambda)$,
 with the observers $y_j$ composing the majority of the 
 central bulk (and with a typical observer describable by 
 $\sim \lambda$ bits).
 
 This is the ``small" case for $\mathcal{U}$, 
 with a finite upper bound for the complexity $I_K$
 of all gauge invariant information structures when 
 expressed in a compact representation 
 (although this ``small" case 
 still corresponds to a very rich and intricate universe 
 of information).
 If this is the nature of $\mathcal{U}$ then presumably as 
 typical nontrivial observers we should 
 already be asymptotically approaching 
 everything that can be known.
 Of course, it is also possible for $\mathcal{U}$ to be finite but 
 for the OCH to be incorrect.
 For instance there could be a very large  
 $\mathcal{Z} \subset \mathcal{U}$ such that 
 all of the elements of $\mathcal{Z}$ are much too 
 complex to be absorbed by any 
 physical observer (and thus 
 $|\mathcal{Z}| \gg |\mathcal{O}| $).
 In this case there must be some 
 special property possessed by the observers, other than 
 being very numerous, so that one should 
 exist as an observer, rather than be an 
 element of $\mathcal{Z}$ 
 (note that we still assume the PCTT and thus everything that 
 exists is some form of information).
 Note that as we do happen to be observers we 
 are effectively ``trapped": we can 
 only point to the possible existence of a 
 vast $\mathcal{Z}$.  
 Any progress towards 
 explicating its various structures would drive 
 $\mathcal{Z}$ to being a subset of $\mathcal{O}$.
 In general we will make the simplifying assumption that 
 these enormous and ``forever inaccessible" $\mathcal{Z}$ 
 sets do not exist.
 
 It would be interesting if $I(\mathcal{U}) < N$, 
 but the alternative is even more intriguing:  
 it may well be that $\mathcal{U}$ is composed  
 of an infinite number of distinct, finitely complex $x_i$.
 We view G\"odel's Incompleteness result 
 (paraphrased as ``No finite 
 system of axioms and inference rules 
 captures all mathematical truths"), 
 and Turing's halting result 
 (``no finite algorithm can capture the behavior of 
 all programs") as suggestive that $I(\mathcal{U}) = \aleph_0$.
 Likewise, the probable result from 
 computational complexity that P $\ne$ NP 
 (and the associated complexity class hierarchy \cite{Aaronson:2004}) 
 is also suggestive that there are nontrivial objects of 
 unbounded complexity in $\mathcal{U}$. 
 However, we do not interpret these results as 
 definitive proof that $\mathcal{U}$ is infinite.
 For instance, repeatedly applying G\"odel's result 
 to generate new independent axioms doesn't actually lead 
 to new structure, and the unpredictability of 
 most programs may be due their being 
 effectively random in structure.
 
 If the OCH is correct, and $I(\mathcal{U}) = \aleph_0$, then  
 the status of observers is more subtle.
 As there are $x_i$ of unbounded complexity in 
 $\mathcal{U}$ there would need to be 
 arbitrarily complex $y_j$ to absorb those 
 structures.  If these $y_j$ do exist the counting 
 would need to be regulated by first considering only 
 $x_i$ with a Kolmogorov complexity of less than 
 $N$ bits, and then letting $N \rightarrow \infty$.
 If the OCH holds, then the density of the 
 observers $|\mathcal{O}|/|\mathcal{U}|$ will 
 approach one as the complexity of structures goes 
 to infinity.
 In this way the observers would roughly resemble the composite 
 integers, with the primes corresponding 
 to the various information structures that they can absorb.
 
 This proposal appears at first to be incompatible 
 with our existence as a typical observers.
 For instance, consider a class $Y_{\alpha} \subset \mathcal{U}$ which is 
 composed of observers one thousand times more complex than a Homo Sapiens: 
 $I_K(y_a) \sim 10^3 I_K(y_b)$, with $y_a \in Y_{\alpha}$ and $y_b \in Y_{HS}$.
 Via combinatorics it follows that $|Y_{\alpha}| \gg |Y_{HS}|$ and thus one should 
 expect to be one of the very advanced $Y_{\alpha}$ 
 instead of a human.
 However, this process can be repeated indefinitely: 
 one should instead expect to be a member of $Y_{\beta}$ 
 whose members are $10^3$ times more complex than 
 those in $Y_{\alpha}$ and so on. 
 As there is no upper bound, 
 there is no one level of complexity that one should expect 
 to exist at -- there will always be many more individuals 
 at much higher levels.
 
 We propose that the resolution for this is that one 
 should expect to exist as a universal observer $\hat{y}$, rather than 
 existing at any one level of complexity.
 A universal observer $\hat{y}$ has the ability to absorb any information 
 structure $x_i$ from $\mathcal{U}$, and they thus 
 collectively dominate the counting among all observers
 (as well as $\mathcal{U}$ itself).
 However, in order to be nontrivial information structures 
 we have argued that the observers must be 
 concretely realized computers embedded in some 
 mathematical structure like $\Psi$.  
 Being concretely realized, an observer will have a 
 finite capacity at any one moment in time, and accordingly there 
 will be an upper bound to the complexity 
 of objects that they can promptly absorb at that time.
 The solution is to not place a time limit 
 for the absorption of a very complex structure: 
 in general a universal observer $\hat{y}_j$ will 
 need to gradually upgrade their physical capacity 
 (via scientific and technological 
 progress) 
 for a long period of time until they have the 
 ability to absorb an arbitrarily large $x_i$ from $\mathcal{U}$.
 Universal observers are thus self similar: 
 at any one moment in time they will find themselves 
 to be more complex than some other individuals, 
 and less complex than a large number of 
 (potential) observers above them.
 Note in particular that they are not structures that exist statically 
 at any one point in time,  
 but rather are entities that are always evolving in time.
 It is because of this capacity for growth that the universal 
 observers -- both concretely realized and continuously evolving  
 --  are the typical observers (if $I(\mathcal{U}) = \aleph_0$).
 
 This scenario seems to be at odds with our current experiences: 
 after reaching adulthood the complexity of what one 
 can learn generally levels off. 
 If this state of affairs was to persist indefinitely, 
 then one should conclude something along the lines of:  
 $\mathcal{U}$ is finite and we are close to 
 saturating what is possible, or: $\mathcal{U}$ 
 is infinite in extent and therefore the OCH is wrong.
 However, there is a real chance that this restriction   
 could be lifted in the coming decades.
 In particular the development of human level 
 artificial intelligence (and then beyond) 
 would open the possibility of unbounded 
 growth.
 We are not yet universal observers, but we may 
 be on the boundary of becoming them.
 %Indeed, it is the peculiar fact that this technological milestone 
 %could take place in the relatively near future 
 %\footnote{Compared to, say, the time scale of hominid evolution.}
 %that has lead us to seriously consider 
 
 Admittedly, the plausibility of the development 
 of human level Artificial Intelligence (AI) is a contentious issue, 
 and there have been promises for its arrival at  
 dates that have long since come and gone \cite{Russell:2002}.
 Still, it can be argued that computing 
 hardware has now become sufficiently advanced 
 to support strong AI. 
 For instance, based on analysis of the 
 retina Moravec \cite{Moravec:2000} estimates the information 
 processing rate of a single neuron to be 
 on the order $10^3$ flops, which results in an 
 estimate of $10^{14}$ flops for the entire brain, which is comparable 
 to the current largest supercomputers. 
 Likewise there is 
 steady progress on the software side in a large 
 number of specialized fronts.
 To give a concrete example, 
 given the advances in 
 robotics, natural language processing, 
 and object recognition, it should now be possible 
 (with sufficient funding) 
 to build a bipedal robot which, 
 if asked, could find and describe 
 various objects while navigating its  
 environment.
 There is still a large distance between 
 this type of robot and an 
 AI that could, say, be a good sitcom writer, but 
 this gulf should be bridgeable by 
 continuous evolutionary improvements.
 The bridging might not even take too long: Kurzweil, for instance, 
 optimistically predicts that strong AI capable of passing 
 the Turing test could arrive by as early as 2030 \cite{Kurzweil:2006}.
 His predictions are based on a coarse grained extrapolation 
 of the growth in information technologies, which so far have followed 
 smooth exponentials for many decades.
 %\footnote{To date quantities such as the numbers of flops 
 %available for a set price have 
 %evolved in a remarkably smooth manner.}. 

 In general, most non-dualists would agree that 
 human level AI is possible in theory, if 
 perhaps disagreeing on a likely arrival date.
 Note also that while the emergence of AI is not assured in any one 
 civilization, that it is effectively guaranteed 
 to occur in some fraction of histories in an 
 eternally inflating universe.
 Furthermore, if it is developed, it would be surprising 
 if it were to then stall permanently at about our intelligence level  
 (although this could be evidence that we are 
 already saturating what is possible in a finite $\mathcal{U}$).
 It seems more likely that they would be able keep 
 on progressing, in time processing thoughts 
 orders of magnitude more complex than 
 we are capable of now.
 %\footnote{One consequence of this would be that 
 %most of their actions would likely be incomprehensible 
 %to regular humans, and thus this transition 
 %has been called a technological singularity.}.
 However, there currently appear to be limits 
 to the complexity of computers that could 
 conceivably be built, which puts the prospect 
 of unbounded growth in doubt \cite{Lloyd:2001bh},\cite{Krauss:2004jy}.
 As noted, Planck scale and holographic 
 arguments indicate that a finite number 
 of degrees of freedom are available 
 for computation in a finite 
 volume of spacetime, and the construction 
 of arbitrarily large structures would at least be a 
 nontrivial undertaking.
 
 These apparent roadblocks are not necessarily 
 insurmountable.
 For instance, Dyson \cite{Dyson:1979zz} argued that 
 life could persist indefinitely in a flat or open universe
 (if in an ever slower, colder form).
 Krauss and Starkman disagree \cite{Krauss:1999hj}, 
 pointing out that with a nonzero $\Lambda$ that 
 temperatures can not be lowered indefinitely 
 (although note \cite{Dyson:2002pf}), 
 and alternatively in a flat or open $\Lambda=0$ universe 
 the amount of matter available to a civilization would 
 either be finite, or would lead to a collapse 
 into a black hole.
 However, they observe that their analysis does not examine 
 speculative strong quantum gravity effects 
 (such as \cite{Borde:1998wa}) which may allow 
 workarounds to these restrictions
 (we also note that true singularities may be 
 avoided in the interiors of black holes - 
 see e.g. \cite{Mathur:2005zp},\cite{Skenderis:2008qn}).
 
 Likewise, the apparent restrictions on the 
 number of degrees of freedom available 
 in a finite volume are not necessarily inviolable.
 These restrictions stem from the current structure  
 of proposed fundamental theories, but these 
 may evolve and generalize in the future.
 To pick an example, the majority of calculations 
 in string theory have been done in the critical 
 dimension $D=10$, but non-critical (linear 
 dilaton) theories are also conceivable \cite{Silverstein:2001xn},\cite{Maloney:2002rr}.
 More abstractly, it has always been the case in the past 
 that the theories we have discovered have been 
 the limiting cases (e.g. $\hbar \rightarrow 0; v/c \rightarrow 0$) 
 of more complex theories, and this may apply 
 to string theory (and thus its implications) as well. 
 Furthermore, if it is possible for an apparently 
 fundamental theory to be a particular limit of 
 a yet more complex structure, 
 then statistically this should turn out to be the case 
 (i.e. there will be many ``deeper structure" 
 variations available in $\mathcal{U}$, 
 as opposed to only one 
 version where the deepest laws are equal to those 
 that have already been discovered).
 
 Finally, the development of powerful AI 
 seems possible within standard atomic physics, 
 and so the subsequent details in the task of unbounded advancement 
 would fall to them (assuming they do arrive).
 Note that a degree of vagueness in these futuristic predictions is 
 unavoidable due to the sparseness of the nontrivial 
 structures $x_i$ among all possible permutations $s_i$
 (compounded by the likely fact that P$\ne$NP).
 Scientific and technological advancement 
 necessarily involves a large amount of trial and error 
 because of this sparseness, 
 and is thus both ``hard" and hard to predict.
 Still, if $\mathcal{U}$ is infinite and the OCH 
 is correct, then in time the AIs will be successful in finding and utilizing  
 modifications of (or loopholes in) basic 
 physics in order to continuously compute 
 and absorb evermore complex information structures, 
 thus becoming universal observers $\hat{y}_j$.  
 %\footnote{Due to the sparseness of the nontrivial 
 %structures among all possible permutations, 
 %scientific and technological advancement 
 %necessarily involves a large amount of trial and error 
 %and is thus ``hard" -- this is likely to always be the case.}.

 %Lastly it is also conceivable that $\mathcal{U} > \aleph_0$, 
 %and there are irreducibly infinitely complex  
 %structures in $\mathcal{U}$.
 %If this was the case and the OCH was also true, 
 %then universal observers would need to be 
 %able to reach ``the limit point at infinity" 
 %(via hypercomputation perhaps)  
 %and proceed onward, absorbing 
 %infinitely complex $x_i$.  
 %In this case one could 
 %presumably switch to cardinality arguments 
 %to describe the size of 
 %$\mathcal{O}$ in $\mathcal{U}$, 
 %instead of the regularized,  
 %``composites among the integers" counting 
 %used for $\mathcal{U} = \aleph_0$.
 %It is possible that $\mathcal{U}$ is this 
 %large, but we find the countable 
 %case to be a good bit more plausible. 
 %Indeed, $\mathcal{U} = \aleph_0$ already 
 %describes an exceptionally vast space, 
 %and is correspondingly speculative.
 %It is the curious fact that human level 
 %AI seems possible in the relatively 
 %near future (with implied subsequent 
 %developments) that encourages us to 
 %seriously consider the possibility that 
 %there is no finite 
 %upper bound to the complexity of real 
 %information structures.

 %%%%%%%%%%%%%%%%%%%%%%%%%%%%%%%%%%%%%%%%%
 %%%%%%%%%%%%%%%%%%%%%%%%%%%%%%%%%%%%%%%%%
 %%%%%%%%%%%%%%%%%%%%%%%%%%%%%%%%%%%%%%%%%
 \section{Application and Predictions}
  
 In his ``Discourse on the Method"
 Ren\'e Descartes searched for the securest 
 foundation for philosophical inquiry, and 
 converged upon ``I think, therefore I am". 
 We disagree with his subsequent claims, 
 but the central observation is quite interesting.
 The insight can be split into two parts: 
 ``I exist" (or just: ``Existence") and then 
 ``I exist as a thinker".
 We assume basic existence 
 (i.e. that there is something rather than nothing),  
 and question the second part: why is it 
 that to exist, to be anything at all, entails 
 existing as a thinker?
 Why is it that to be an observer is the 
 particular form that existence takes?
 
 We have proposed a statistical answer in the OCH.
 The first clue emerges from the 
 great unity in the world that the scientific 
 revolution has revealed in the last 350 years.
 %:the link between the falling apple and the 
 %motion of the planets (and thus between math and reality), 
 %the common origin of all life on Earth, 
 %to the derivation of light from 
 %electromagnetism and thermodynamics from 
 %statistical mechanics, and the subtle, 
 %nonintuitive union of space and time itself.
 While many of the details remain to be filled in, 
 the broad sweep of phenomena that we observe  
 fits neatly into an overarching explanatory structure, 
 from mathematics and particle physics 
 to atomic physics and chemistry, up through 
 microbiology and neurology to consciousness 
 and the evolution of civilization.  
 %The great range of scales in the universe is also impressive, 
 %from the interactions between quarks and 
 %gluons in a single proton to the cosmic 
 %microwave background spread across the sky
 %(and this is only a small sample of what exists  
 %if the Many Worlds Interpretation or eternal 
 %inflation are correct).
 %Likewise we note the great diversity 
 %present in a single mind, from all of these 
 %types of ideas to the various sensory qualia and the sense of self, 
 %all of which stemming from the patterns generated by 
 %interacting neurons in the brain.
 %The amount of diversity possible in a single 
 %mind is also striking -- from 
 It is striking that all of the objects in this expansive range 
 can be fully described as various types 
 of information  
 (assuming the physical version of the Church Turing Thesis is correct).
 The universal capacity of information for representation 
 is likewise reflected in the way in which all 
 the elements of conscious experience (from any type of idea 
 to memories, emotions, sensory qualia, and the sense of self) 
 are equal to the different patterns created by interacting neurons.
 
 Following this observation it is natural to postulate 
 a ``grand unification": 
 everything that exists is a type of information.
 The $x_i$ can be collected in a $\mathcal{U}$, 
 which removes gauge redundancies and random 
 strings.
 The observers then form a subset of the grand ensemble 
 with the special property that they 
 selectively extract information from other sources. 
 If the Observer Class Hypothesis is correct, 
 the combinatorics that arise from this selection 
 process lead to the observers dominating the 
 counting among all information structures (\ref{ob_dom}).
 This provides a statistical explanation for 
 why we find ourselves to be thinkers:  
 it is by far the most probable form of existence.

 We have considered two specializations of the hypothesis:
 in one version $I(\mathcal{U}) < N$ 
 there is a finite upper bound to the complexity 
 of objects $x_i$ (when expressed in a compact representation),
 and in the other $I(\mathcal{U}) = \aleph_0$ there is no upper bound.
 The first possibility could be viewed as the null assumption -- 
 currently the complexity of the universe looks finite.
 However, we will adopt the more extravagant second 
 version in order to apply the OCH and make predictions.
 This is version is at least conceivable: for instance both G\"odel and Turing's results 
 are suggestive that there is no upper bound to the complexity 
 of real information structures.
 If $I(\mathcal{U}) = \aleph_0$ then the OCH requires the existence of 
 universal observers $\hat{y}_j$ which can 
 absorb any information structure. 
 Being physically realized, 
 they will in general need to upgrade their capacity 
 for a long period of time before they have 
 the ability to absorb a particular $x_i$.  
 The universal observers 
 are thus concretely realized, continuously evolving in time, 
 and broadly self-similar.
 The curious fact that strong AI could be developed in the coming decades  
 is supportive of the possibility that 
 universal observers could be developed.
 
We can now apply the universal observer version of 
the OCH to several problems involving typical observers.  
We first examine the Doomsday argument 
due to Carter \cite{Carter:1983} and Gott \cite{Gott:1993}.
In the simplified version one considers 2 possible Future 
Histories: a ``utopia" FH1, and a ``doomsday" version FH2.
In FH1 our species perseveres for a very long 
time (say $\sim 10^8$ years), so that perhaps $\sim 10^{16}$ 
humans will live at some point.
In the alternative FH2, some terrible 
event eliminates our species in the fairly near future 
(say within the 21st century) so that the total number 
of humans that will ever live is comparable to the number 
that have been born so far (i.e. from $10^{10}$ to $10^{11}$).
Note that we are typical observers in the doomsday history FH2, 
but are very special observers in the FH1 scenario: in the utopian 
version we live very close to the beginning of history.  
If the prior probabilities for FH1 and FH2 are comparable 
(a large assumption), then using Bayesian statistics 
one would conclude that the disastrous FH2 scenario 
will likely take place.

The conclusion is flipped if we assume that we are 
universal observers however.
This is because the observations of $\hat{y}_j$ are roughly self similar: 
at any one moment in time they can promptly absorb 
$x_i$ of some finite complexity, but at arbitrarily distant points in the future 
they will be able to absorb arbitrarily more complex structures.
All universal observers therefore find themselves to 
be relatively close to the beginning of history.

We can also address the peculiar Boltzmann Brains (BBs)
which seem to affect the choice of measure in cosmology. 
We argue that the assumptions of $I(\mathcal{U}) = \aleph_0$ 
and universal observers are not needed here: in a finite 
$\mathcal{U}$ one also should expect to not be a BB, even if they 
are generated late in the evolution of 
$\Psi$ type universes.
The real information content in $\Psi$ is contained in the 
early non-equilibrium period where a certain subset of patterns 
are produced much more frequently (because of the details 
of their internal structure) than a direct counting of all 
permutations $\Omega$ would indicate.
These nontrivial patterns form the information content 
of $\Psi$, and they are the objects that one could 
exist as.
The Boltzmann Brains are random fluctuations 
that happen to reproduce the patterns that exist in 
the stellar evolution phase -- they carry no information content independent 
of the nontrivial observers that they copy.
If one could exist as a thermal fluctuation, 
then one would be a completely random arrangement, 
not one of the BBs which comprise a tiny fraction of the 
permutation space.
Thermal fluctuations and random structures in 
general contain little real information (despite first appearances) 
as they can be efficiently encapsulated as 
trivial $r_i$ (i.e. they have very low effective complexity \cite{Gellmann:1996}).
Note that if universal observers are possible then the case 
against BBs is further strengthened: 
one should exist as a constantly evolving and 
growing observer, while the BBs 
cease to exist only moments after they are formed.

The concept of Boltzmann Brains has recently been used as 
a tool in the construction of measures in cosmology, 
and to argue that vacuum energy must decay 
on a fairly fast timescale.
We in effect argue that this particular tool is not 
valid -- one should exist as a real, 
gauge-invariant information structure, such as the 
ordinary observers that can emerge after several 
billion years of evolution.
Anthropic arguments and measure proposals 
may still be useful, but they need to be predicated 
on these nontrivial observers.
This is the case for instance in Weinberg's 
prediction for the cosmological constant, 
or in Dicke \cite{Dicke:1961} and Carter's \cite{Carter:1974} explanations of Dirac's large 
number coincidences.

The OCH can also be used to make coarse-grained predictions
about the evolution of minds and civilization going into the future.
Previously we have considered the possibility 
of the development of human level AI (and then beyond) 
as supportive of the large  
$\mathcal{U}$ version of the OCH.
We switch here and assume $I(\mathcal{U}) = \aleph_0$, 
and thus the typical observer (and therefore the typical form of 
information) is a universal observer $\hat{y}_j$.
This then necessitates the development of strong AI, 
to be followed by ceaseless advancement and exploration of ever more 
complex $x_i$.
Note that the universal observer version of the OCH probably can not be  
definitively proven as it hinges upon the existence of arbitrarily complex nontrivial 
structures.
Rather, in practical terms, the longer that scientific and technological 
progress persist the more confidence can be had that it is 
correct.
The current prediction is thus limited to the 
arrival of strong AI, as the detailed prospects for 
subsequent progress would then fall to them.
A precise date for the emergence of strong AI can not 
be given, but it is striking that the timescale appears to be only on the 
order of decades (which compares favorably, 
for instance, to the $\sim 10^{10}$ year  
vacuum decay predictions).

Finally we emphasize the explanatory power of the 
Observer Class Hypothesis.
It is easy to take existing as an observer for granted, 
but upon reflection there should be something special about minds 
so that to exist at all is to exist as one of them.
The OCH answers in both ways: observers are not 
special as they are just another type of 
information, but they are special since their ability 
to absorb other forms of information makes them 
very numerous.
%To be an observer is thus by far the likeliest form 
%of existence.
Further consideration points towards only counting 
over distinct, gauge-invariant forms of information.
One thus needs nontrivial observers, 
and these can emerge from biological evolution 
working within the constraints imposed by 
a mathematical structure like $\Psi$.
If there is no upper bound to the complexity 
of information structures, then existing as a entity 
that is continuously evolving in time is also natural: this 
solves the dual constraints of being physically 
realized while making possible the absorption of 
any $x_i$.
The most common form of information, 
and most probable form 
of existence, is thus  
a typical observer who is embedded in a structure like 
$\Psi$ and is continuously evolving in time. 
%-- time will tell if this is correct.

%Note that the Boltzmann Brains are patterns formed out
%of arrangements of atoms, as is everything else 
%generated 

%================================BIBLIOGRAPHY=====================================

 %The over counting problem has a direct 
 %solution, which also resolves the apparent 
 %numerical dominance of noise states.

%\bibliographystyle{unsrt}
%\bibliographystyle{h-physrev}
%\bibliographystyle{h-physrev5}
%\bibliographystyle{h-elsevier3}
%\bibliographystyle{hep}
%\bibliographystyle{plain}
%\bibliographystyle{kp}
%\bibliography{ochbib}

\end{document}